# High-speed and single-mode FP laser based on parity–time symmetry

SIKANG YANG,[1,3] JING LUAN,[1,3] YU HAN,[1,3] RUIGANG ZHANG,[1,3] QI TIAN,[1,3] PENGXIANG HE,[1,3] DEMING LIU,[1,3] AND MINMING ZHANG[1,2,3,4*]

[1]*School of Optical and Electronic Information, Huazhong University of Science and Technology, Wuhan 430074, China*
[2]*Wuhan National Laboratory for Optoelectronics, Wuhan 430074, China*
[3]*National Engineering Laboratory for Next Generation Internet Access System, Wuhan 430074, China*
[4]*Optics Valley Laboratory, Wuhan 430074, China*
*mmz@hust.edu.cn

**Abstract:** The ability to manipulate cavity resonant modes is of critical importance in laser physics and applications. By exploiting the parity time (PT) symmetry, we propose and experimentally realize a single-mode FP laser with improved output power and high-speed modulation have been demonstrated. The proposed PT symmetric laser consists of two coupled structurally identical FP resonators. The gain and loss in two FP resonators can be manipulated independently by changing the injection currents. In the PT symmetric FP laser, single-mode operation is accomplished by selectively breaking of PT symmetry depending solely on the relation between gain-loss and coupling. Single-mode lasing with output power of 1.7 dBm and a sidemode suppression ratio (SMSR) exceeding 24 dB is demonstrated. The 3 dB bandwidth of 7.9 GHz is achieved and clear eye-openings were obtained for 2.5 Gbps and 10Gbps NRZ operation over 10 km single-mode fibers. Furthermore, the PT symmetry breaking is experimentally confirmed with measured loss and coupling coefficient of two FP resonators. The influence of cavity length, facet reflectivity, and electrical isolation between two P-side electrodes on the side mode suppression ratio and output optical power is also been demonstrated, paving the way for further improvement of the PT symmetric FP laser.

## 1. Introduction

Fabry–Pérot (FP) cavity is one of the typical components of photonic integrated circuit, which is also the ideal candidate for integrated semiconductor lasers for their small footprint and easy to fabrication. Unfortunately, FP lasers support multiple longitudinal resonances within the broad gain bandwidth of semiconductors, which hinders its application especially in wavelength division multiplexing communication systems. Consequently, additional approaches must be employed for single-mode operation. At present, most approaches rely on the use of intra-cavity optical feedback, such as distributed Bragg mirrors [1,2] and distributed feedback gratings [3-5], or extreme confinement of light in subwavelength structures using metallic cavities [6-8]. However, these schemes introduce further demands in terms of fabrication complexity.

Lately, parity time symmetry has raised considerable attention in photonics [9-12]. PT symmetry was firstly developed within in quantum mechanics. In general, a nonconservative optical system with balanced gain-loss distribution is considered an ideal platform for studying the fundamentals of PT symmetry. As indicated in a number of studies, gain and loss can be integrated as nonconservative ingredients to create new effects such as coherent perfect absorption [13-16], unidirectional invisibility [17-19], chiral mode conversion [20-23], sensitivity enhancement at an exceptional point [24-26] and so on. Explorations of parity-time symmetry also offers a unique pathway for mode manipulation in lasers [27-31]. Several promising experiments show that PT symmetry breaking can be employed to establish single-mode operation in inherently multimoded microring lasers with optical pumping [27-29]. An electrically pumped integrated microring laser that supports single-mode operation based on PT symmetry with output power of -14dBm has also been demonstrated [30]. However, for optical communication application, a higher output power laser is needed.

In this letter, we propose and experimentally demonstrate an electrically pumped FP semiconductor laser that supports single-mode lasing with improved output power based on parity–time symmetry. The PT symmetric laser consists of two structurally identical FP resonators. By incorporating two isolated electrodes in each of the two FP resonators, the gain–loss can be controlled independently by changing the injection currents. In the PT symmetric FP laser, single-mode operation is accomplished by selectively breaking of PT symmetry depending solely on the relation between gain-loss and coupling. Single-mode lasing with output power of 1.7 dBm and a sidemode suppression ratio exceeding 24 dB is experimentally demonstrated. The 3 dB bandwidth of 7.9 GHz is achieved and clear eye-openings are obtained for 2.5 Gbps and 10 Gbps NRZ operation over 10km single-mode fibers.

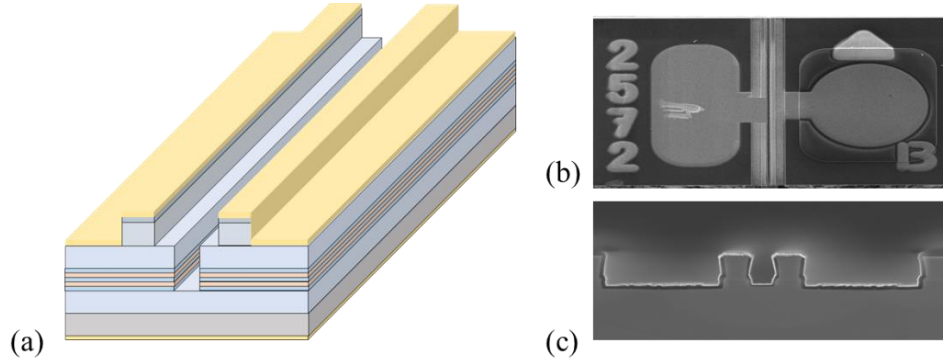

Fig. 1. (a) Schematic structure of the proposed electrically pumped single mode PT-symmetric FP semiconductor laser, (b) and (c) Scanning electron-microscope images of the single mode PT symmetric FP laser from the top and front view respectively.

## 2. Theoretical designs

Fig. 1(a) shows the schematic structure of the electrically pumped single mode PT-symmetric FP semiconductor laser, which consists of two coupled structurally identical ridge waveguide. To establish a PT-symmetric photonic system that the real and imaginary parts exhibit even and odd spatial distributions respectively, two isolated P-side electrodes are employed. In this way, the gain–loss can be controlled independently by changing the injection currents. For sufficient electrical isolation, a deep etching scheme is adopted as shown in Fig. 1(a). The laser is designed around 1330nm with waveguide width of 2.5μm and gap around 2.5μm for suitable coupling coefficient between two ridge waveguides.

According to the coupled mode theory, the evolution of the modal in two coupled ridge waveguides is described as below:

$$\frac{da_m}{dz} = i\beta_m a_m + i\kappa_m b_m + \gamma_a a_m$$

$$\frac{db_m}{dz} = i\beta_m b_m + i\kappa_m a_m + \gamma_b b_m$$

Where $a_m$ and $b_m$ are the amplitude of m'th modes in these two arrays, $\beta_m$ is the propagation constant, $\kappa_m$ is the coupling coefficient between the two ridge waveguides and $\gamma_a$ and $\gamma_b$ are the modal gain and loss of these two ridge waveguides respectively. In this way, the propagation constants of the PT-symmetric laser arrays can be given as:

$$\beta_m = \beta_{m0} + i\frac{\gamma_a + \gamma_b}{2} \pm \sqrt{\kappa_m^2 - \left(\frac{\gamma_a - \gamma_b}{2}\right)^2}$$

Fig. 2(a) and (b) profile the real and imaginary parts of eigenvalues evolution curves versus the loss in the loss ridge waveguide based on Eq. (2). When $(\gamma_a - \gamma_b)/2 < \kappa_m$, the real parts are split, and the imaginary parts are consistent, implying that two eigenvalues of this non-Hermitian system can be entirely real-valued with a frequency difference, and all modes experience the same gain or loss. However, when $(\gamma_a - \gamma_b)/2 > \kappa_m$, the real parts are coalescent, and the imaginary parts are split, indicating that the PT symmetry was spontaneously broken, and a conjugate pair of lasing and decaying modes emerges. Therefore, with appropriate gain-loss contrast and coupling coefficient, single-mode operation is accomplished by selectively breaking of PT symmetry that only the mode with the highest gain reaches PT symmetry breaking threshold while the other modes keep in the PT symmetry phase, as shown in Fig. 2(b). The Finite Difference Eigenmode (FDE) solver of MODE Solution was used to evaluate its performance. As shown in the insets in Fig. 2(c), in the single mode region, only the mode with the highest gain above PT symmetry breaking threshold, and the modes predominantly reside mostly in one of the ridges, and consequently experience a net amount of gain or loss. In contrast, the other modes of FP laser are below the PT symmetry breaking threshold, and the supermodes are evenly distributed between the gain and loss ridges and consequently experience neutral oscillations.

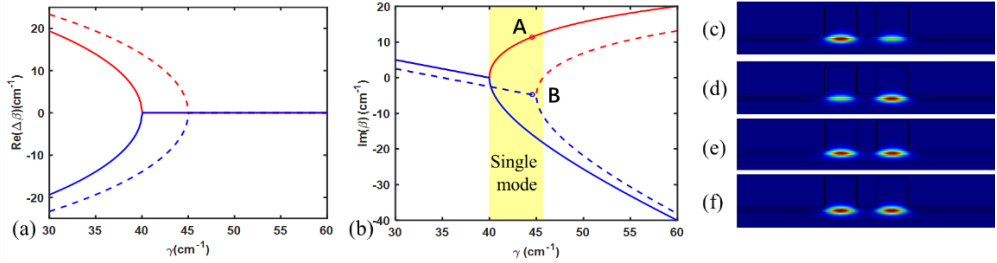

Fig. 2. (a) and (b) Simulated real and imaginary part of the eigenfrequency versus loss in the loss ridge waveguide respectively, (c) and (d) simulated electric field of two supermodes at parity time symmetry broken phase (Point A in Fig. 2. (b)), (e) and (f) simulated electric field of two supermodes at parity time symmetry phase (Point B in Fig. 2. (b))

## 3. Experiment results

The single mode PT symmetric FP laser was experimentally demonstrated using standard nanofabrication techniques. The FP laser heterostructure is grown by metal organic chemical vapor deposition (MOCVD). The coupled ridge waveguides are defined by standard photolithography process and etch processes of inductively coupled plasma etching and wet etching. The metal stacks are deposited on the wafer by magnetron sputtering to form P-side and N-side electrode. Fig. 1(c) shows the SEM images of the single mode PT symmetric FP laser from the top and front view. In the whole fabrication process, no regrowth step and electron-beam lithography are required, which greatly reduces the complexity and cost of fabrication.

A 150 μm long laser with ridges width of 2.5 μm, gap of 3.6 μm, front facet reflectivity of 30% and rear reflectivity of 92% is tested for optical spectra and light-current characterization. An optical spectrum analyzer (Yokogawa AQ6370C-20) with resolution of 0.1nm are utilized to measure the optical spectra of the laser. As shown in Fig. 3(b), an emission spectrum with multiple modes in a single ridge FP resonator is observed. For the double ridge FP laser, an emission spectrum with multiple modes is observed and the coupling induced mode splitting can be seen when two ridge waveguides are electrically even pumped with the same currents, as shown in Fig. 3(d). Once the PT symmetry is broken by tuning the injection currents of two ridge waveguides, single-mode lasing with a SMSR of 24 dB at 1329 nm occurs as shown in Fig. 3(c), with the injection currents of 22.5mA and 0mA. The characteristic light-current curves for three arrangements of evenly pumped double ridge, PT symmetric and single ridge is shown in Fig. 3(a). The current is the sum of the injection currents of the two electrodes and the light from the laser is coupled out of the chip using a lensed fiber. The PT-symmetric arrangement appears greater slope efficiency and smaller threshold current than those of evenly pumped arrangement, indicating that the presence of the lossy ridge waveguide only serves to suppress the unwanted modes without necessarily decrease the overall efficiency. The greater slope efficiency of PT-symmetric arrangement is partly due to the higher coupling efficiency with the optical lensed fiber and smaller intracavity loss, and the output power of 1.7 dBm is accomplished. The near and far field patterns of the evenly pumped double ridge laser, PT symmetric laser and single ridge laser are collected by the 100X objective lens and near infrared CCD camera as shown in Fig. 3(e)-(j). The nearfield is evenly distributed between the gain and loss ridges when the double ridge is evenly pumped and the far field is double lobes, which is consistent with the simulation result for lasers bellowing the PT symmetry breaking condition. Once the PT symmetry is broken with the injection currents of 22.5mA and 0mA, the near field of laser predominantly reside in the ridge of gain and single-lobe far-field pattern is obtained, which almost the same as in the single ridge laser.

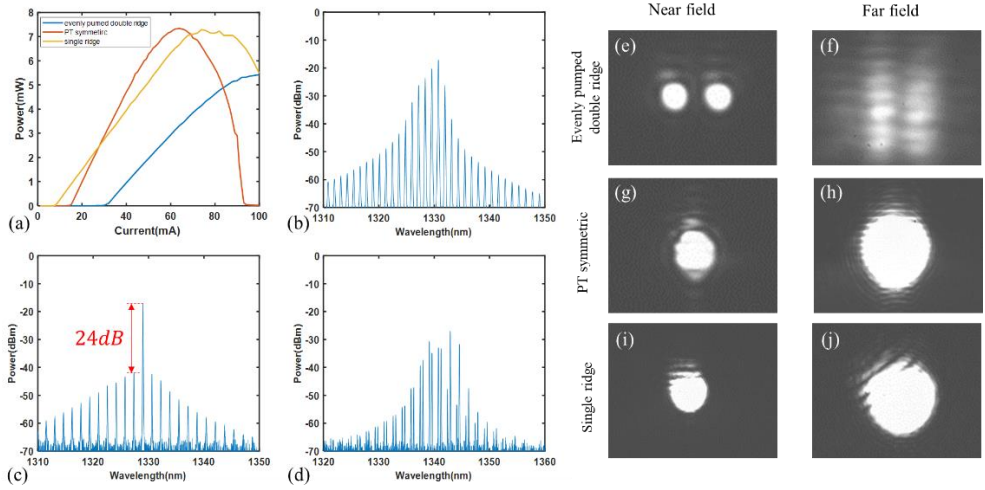

Fig. 3. (a) Characteristic light-current curves for three arrangements of evenly pumped double ridge, PT symmetric and single ridge, (b)-(d) optical spectrum under three arrangements of evenly pumped double ridge, PT symmetric and single ridge respectively, and (e)-(j) Near- and far-field patterns under three arrangements of evenly pumped double ridge, PT symmetric and single ridge respectively.

Furthermore, the PT symmetric phase transition at single mode lasing region is experimentally confirmed with measured internal loss $\alpha_i$ and coupling coefficient κ of two FP resonators. The internal loss of FP cavity can be characterized from the below-threshold amplified spontaneous emission (ASE) spectrum of FP lasers with Fourier series expansion (FSE) method. The below-threshold ASE spectrum of the FP laser is shown in the Fig. 4(a). The internal loss of FP cavity is about 23 $cm^{-1}$ as shown in Fig. 4(b), and the more detailed procedures can be found in [32]. The coupling coefficient between the two FP resonator is exacted experimentally from the emission spectra under uniform pumping condition [33]. The propagation constant splitting δβ of two uniform pumped FP resonator is directly proportional to the coupling coefficient, δβ = 2κ. Therefore, the resonant wavelength splitting δλ of two uniform pumped FP resonator can be related to the coupling coefficient, $κ = πn_{eff}Δλ/λ^2$. As shown in Fig. 3(a), the resonant wavelength splitting δλ = 0.48nm, indicates the coupling coefficient κ = 31 $cm^{-1}$. Finally, the net mode gain g = 43 $cm^{-1}$, is calculated by lasing threshold condition that the net modal gain $g = Γg_0 − α_i$ needs to compensate for mirror loss $α_m$. In this way, the PT symmetry breaking condition $(γ_a − γ_b)/2 > κ_m$ is experimentally confirmed.

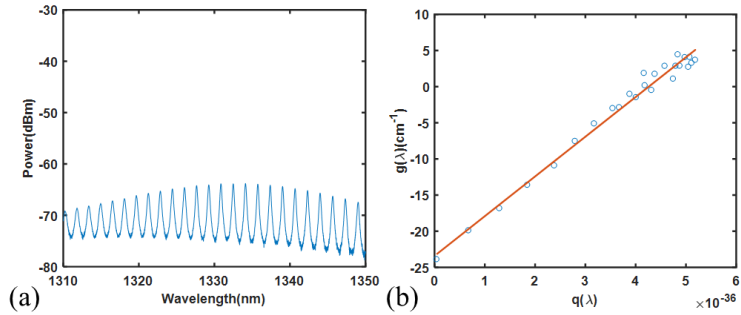

Fig. 4. (a) Amplified spontaneous emission spectrum of FP lasers the below-threshold, (b) internal loss of FP cavity characterized with Fourier series expansion method

Fig. 5. (a) shows the small-signal response of the evenly pumped double ridge laser with the injection currents of 22.5mA and 22.5mA, PT symmetric laser with the injection currents of 22.5mA and 0mA and single ridge laser with the injection currents of 22.5mA. The 3dB bandwidth of PT symmetric laser is 7.9GHz, which is a little lower than the single ridge laser for the little difference of threshold current. Evenly pumped double ridge laser shows the lowest relaxation oscillation frequency due to the large cavity loss. With the increases of injection currents, the PT symmetric laser demonstrate the same 3dB bandwidth with the single ridge laser, indicating that PT symmetry breaking have little impact on small-signal response of lasers as shown in Fig. 5. (d).

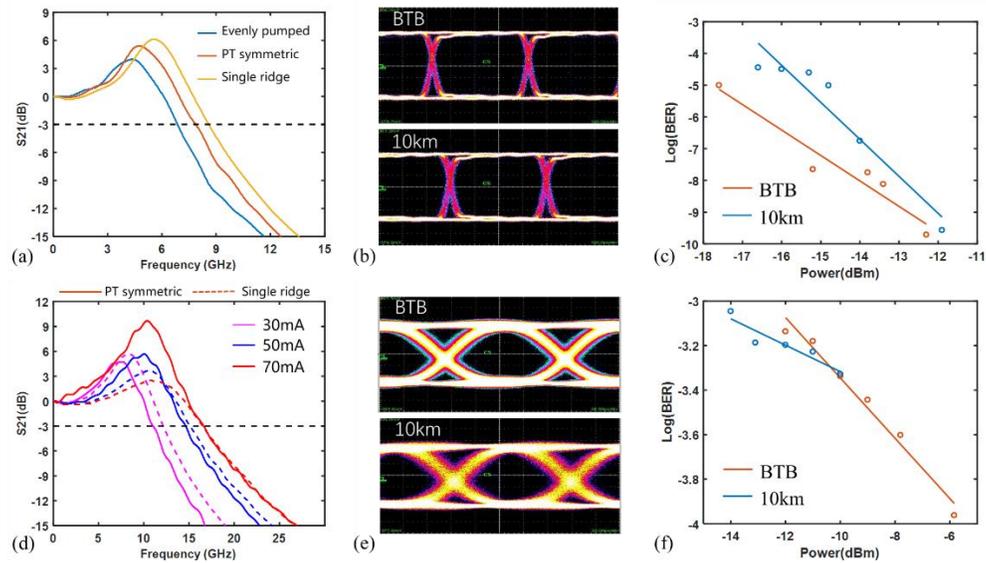

Fig. 5. (a) small-signal response for three arrangements of evenly pumped double ridge, PT symmetric and single ridge, (d) small-signal response for different injection currents, (b) and (e) eye diagrams in BTB and 10km SMF transmission configuration for 2.5 Gbps and 10 Gbps NRZ signal respectively, (c) and (f) BERs characteristics as a function of received optical power in both BTB and 10km SMF transmission configuration for 2.5 Gbps and 10 Gbps NRZ signal respectively

We also directly modulate the PT symmetric FP laser with a 2.5 Gbps NRZ signal in both back-to-back (BTB) and 10km single mode fiber (SMF) transmission configuration. A clear eye opening is obtained at a bias current of 22.5 mA and 0mA as shown in Fig. 5. (b). The measurement of the bit-error rate (BER) for 2.5 Gbps signals in both BTB and 10km SMF transmission configuration is also performed. The bias current of the PT symmetric FP laser is set to 22.5 mA and 0 mA, the same with single-mode lasing condition. Fig. 5. (c) shows BERs characteristics as a function of received optical power in both BTB and 10km SMF transmission configuration. The BER after the 10 km transmission was less than in the BTB configuration due to the dispersion effect of the SMF, and near error-free operation is still obtained for 10-km SMFs transmission. Furthermore, we expand the bit rate up to 10 Gbps NRZ at the same bias current. Clear eye opening is also obtained for both BTB and 10km SMF transmission configuration, and there is some degradation in BERs characteristics compared with 2.5 Gbps signals as shown in Fig. 5. (e) and (f). These results indicate the PT symmetric FP laser can be a great candidate for high speed optical communications.

## 4. Discussion

To further investigate the influence of cavity length, facet reflectivity, and electrical isolation between two P-side electrodes on the side mode suppression ratio and output optical power, four kinds of PT symmetric FP laser with different structures are fabricated. As shown in Fig. 6(a), the average SMSR is 14 dB for the 200 μm long PT symmetric FP lasers with front facet reflectivity of 30% and rear reflectivity of 92%, and 10dB for front facet reflectivity of 30% and rear reflectivity of 30%. The better performance of higher facet reflectivity is due to the higher quality factor. As shown in Fig. 6(d), the PT symmetric FP lasers with higher facet reflectivity also show higher output power due to smaller mirror loss. PT symmetric FP lasers with shorter cavity length have also been demonstrated. As shown in Fig. 6(b), lasers with shorter cavity length show a slight improvement for SMSR. The measured free spectral ranges (FSR) are 1.18 nm and 1.59 nm for laser with cavity length of 200 μm and 150 μm respectively. The small FSR increasement contribute to the slight improvement of SMSR. Finally, two kinds of electrical isolation are considered. The first one is accomplished just by two separated P-side electrodes, and the measured resistance is only 1 kΩ. The second method is deep etching as shown in Fig. 1(a) and the measured resistance reaches 60 kΩ. As shown Fig. 6(a), the average SMSR is 15 dB and 18 dB for the first and second method respectively. Deep etching demonstrates some advantage in terms of high SMSR but at the sacrifice of output power for large cavity loss.

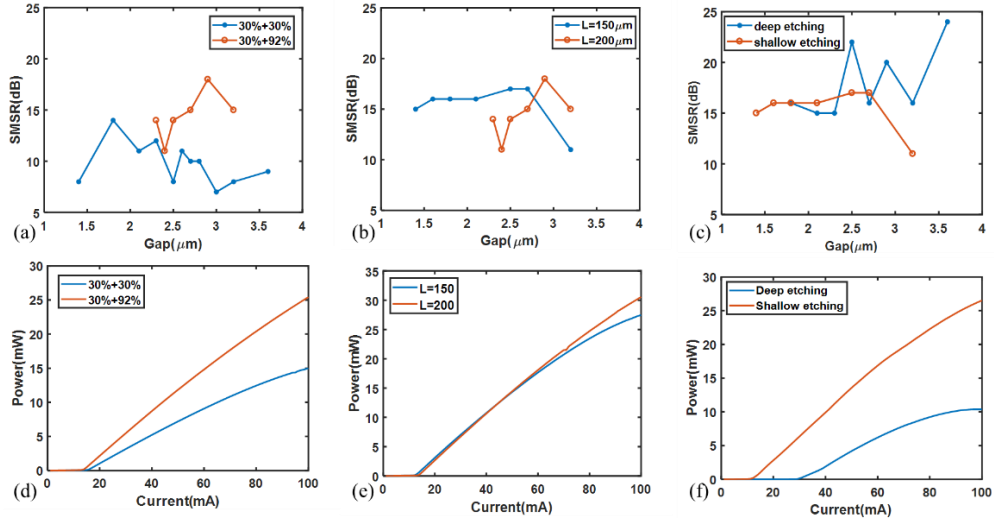

Fig. 6. (a)-(c) Side mode suppression ratio, and (d-f) characteristic light-current curves for PT symmetric FP lasers with different facet reflectivity, cavity length, and electrical isolation respectively.

## 5. Conclusion

As a conclusion, we propose and experimentally demonstrate an electrically pumped FP semiconductor laser that supports single-mode lasing with an improved output power based on parity–time symmetry. Single-mode lasing with output power of 1.7 dBm and a sidemode suppression ratio exceeding 24 dB is experimentally demonstrated. The PT symmetry breaking condition is also experimentally confirmed with measured loss and coupling coefficient of two FP resonators. Higher sidemode suppression ratio can be realized with shorter cavity length, higher facet reflectivity and greater electrical isolation. The 3dB bandwidth of 7.9GHz is achieved and clear eye-openings were obtained for 2.5Gbps and 10Gbps NRZ operation over 10km single-mode fibers. The results demonstrate that the coupled FP resonators can be a great candidate for high speed optical communications.

**Funding**

This work is supported by the National Key Research and Development Program of China (2018YFB2201500).